\documentclass[12pt]{article}

\begin{document}

\title{\bf The ``proton spin crisis" -\\a quantum query}

\date{}

\author{Johan Hansson \\ Department of Physics \\ Lule{\aa} University of Technology \\ SE-971
87 Lule\aa, Sweden}

\maketitle

\begin{abstract}
The ``proton spin crisis" was introduced in the late 1980s, when
the EMC-experiment revealed that little or nothing of a proton's
spin seemed to be carried by its quarks. The main objective of
this paper is to point out that it is wrong to assume that the
proton spin, measured by completely different experimental setups,
should be the same in all circumstances.
%\\
%PACS numbers: 03.65.-w, 03.65.Bz, 11.15.-q
\end{abstract}

%%\newpage

%%\pacs{ PACS numbers: 03.65.-w, 03.65.Bz, 11.15.-q }
The ``proton spin crisis"\cite{Anselmino} essentially refers to
the experimental finding that very little of the spin of a proton
seems to be carried by the quarks from which it is supposedly
built. This was a very curious and unexpected experimental result
of the European Muon Collaboration, EMC \cite{EMC} (later
confirmed by other experiments), as the whole idea of the original
quark model of Gell-Mann \cite{Gell-Mann} and Zweig \cite{Zweig}
was to account for 100 percent of the hadronic spins, solely in
terms of quarks. This original, or ``naive", quark model also was
very successful in explaining and predicting hadron spectroscopy
data.

The purpose of this paper is to point out that the ``proton spin
crisis" may be due to a misinterpretation of the underlying,
quantum mechanical theory. As spin is a fundamentally quantum
mechanical entity, without any classical analog, special care must
be taken to treat it in a correct quantum mechanical manner.

According to Niels Bohr, \textit{the whole experimental setup}
must be considered when we observe quantum mechanical systems. It
means that a quantal object does not ``really exist" independently
of how it is observed. This notion was later quantified by Bell
\cite{Bell}, and verified experimentally by Clauser and Freedman
\cite{Clauser}, Aspect, Dalibard and Roger \cite{Aspect} and
others. These experimentally observed violations of Bell's theorem
\cite{Bell} are in accordance with quantum mechanics, but
incompatible with a locally realistic world view, meaning that
quantum objects do not have objective properties unless and until
they are actually measured\footnote{To be exact, also the
possibility exists of non-local ``hidden variable" theories, where
objects \textit{do} exist at all times. However, such theories
manifestly break Lorentz-covariance.}. The quantum states are not
merely unknown, but completely undecided until measured. It is
important to stress that \textit{this is not merely a
philosophical question, but an experimentally verified prediction
based upon the very foundations of quantum theory itself}. To
quote John Wheeler: ``No elementary quantum phenomenon is a
phenomenon until it is a registered (observed) phenomenon"
\cite{Wheeler}. Unless a specific observable is actually measured,
it really does not exist. This means that we should not \textit{a
priori} assume that different ways of probing the system will give
the same results, as the \textit{system itself} will change when
we change the method of observation.

For the spin of the proton, let us compare two different
experimental setups designed to measure it:

i) The Stern-Gerlach (S-G) experiment, which uses an inhomogeneous
magnetic field to measure the proton spin.

ii) Deep inelastic scattering (DIS), which uses an elementary
probe (electron or neutrino) that inelastically scatters off the
``proton" (actually elastically off partons).

We should at once recognize i) and ii) as different, or
complementary, physical setups. If one measures the first, the
other cannot be measured simultaneously, and vice versa. DIS
disintegrates the proton and produces ``jets" of, often heavier,
hadrons as the collision energy is much larger than the binding
energy, so there is no proton left to measure. Also, the very fact
that the hard reaction in DIS is describable in perturbation
theory means that we are dealing with a different quantum
mechanical object than an undisturbed proton.

In the case of using a S-G apparatus to measure the spin, the
proton is intact both before and after the measurement, potential
scattering being by definition elastic. S-G thus measures the
total spin of the proton, but does not resolve any partons. It
therefore seems natural to identify the spin of an undisturbed
proton with the result from a Stern-Gerlach type of experiment.

As i) and ii) do not refer to the same physical system, the ``spin
sum-rule", usually taken to be an equality, instead reads
\begin{equation}
\frac{\Sigma}{2} + L_q + L_G + \Delta G \neq \frac{1}{2},
\end{equation}
as the left hand side describes the measured spin of the partons,
while the right hand side describes the spin of the proton. The
quantities above stand for: $\Sigma $ = fraction of proton's spin
carried by the spin of quarks and anti-quarks, $L_q $ = quark
orbital angular momentum contribution, $L_G $ = gluon orbital
angular momentum contribution, $\Delta G$ = gluon spin
contribution.

Now, we can try to disregard the fundamental insight from quantum
mechanics described above, and ``force" the proton to always be
described in terms of ``clothed" partons, the so-called
constituent quarks of the naive quark model. The canonical
``minimal coupling" substitution for including interactions
\begin{equation}
p_{\mu} \rightarrow p'_{\mu}  = p_{\mu} + g \, T_a A^a_{\mu},
\end{equation}
implies that the quark and color fields become irrevocably admixed
in an undisturbed proton. Here $p_{\mu}$ is the four-momentum of a
hypothetical free quark, $g$ is the coupling constant of quantum
chromodynamics (QCD) and $T_a A^a_{\mu}$ the combined contribution
of the color (``gluon") fields. If taken literally, it means that
hadrons may be treated as being composed of ``hybrid particles"
with four-momentum $ p'_{\mu}$. An attempt to treat the spin of an
undisturbed proton in terms of such quasi-particles is being
investigated in \cite{Turk}.

An additional complication is the following: While in quantum
electrodynamics (QED) an atomic wave function can approximately be
separated into independent parts due to the weak interaction, and
the spins of the constituents (nuclei and electrons) can be
measured separately as they can be studied in
isolation\footnote{Wigner's classification of particles according
to their mass and spin is given by irreducible representations of
the Poincar\'{e} group, \textit{i.e.} noninteracting fields.}, in
QCD it fails as the interactions between fields in an undisturbed
proton are much stronger than in the QED case, making even an
approximate separation impossible. Even worse, in QCD at low
momentum transfers\footnote{More precisely, the elementary quanta
of QCD are defined only as the momentum transfer goes to infinity.
}, like in an undisturbed proton, the particles ``quarks" and
``gluons" cannot even be defined \cite{Hansson} and thus do not
``exist" within the proton, even when disregarding the quantum
mechanical measurement process described above.

Even if we assume that (``clothed") partons within the proton
\textit{are} defined and approximately obey the Schr\"{o}dinger
equation, the proton spin wave function, $|\chi\rangle$, cannot be
factorized into its separate quark spin wave functions
($|\chi_1\rangle, |\chi_2 \rangle, |\chi_3 \rangle$) as it would
not be an eigenstate of the strongly spin-dependent Hamiltonian,
\begin{equation}
| \chi \rangle= |\chi_1 ,\chi_2 , \chi_3 \rangle \neq
|\chi_1\rangle |\chi_2 \rangle |\chi_3 \rangle .
\end{equation}
In reality the particles are always correlated and the wave
function can never be separated into product states, except as an
approximation when the interaction is sufficiently small.
Actually, the total proton wave function would be
\begin{equation}
\Psi (\mathbf{x}_1, \mathbf{x}_2, \mathbf{x}_3, s_1, s_2, s_3)
\neq u(\mathbf{x}_1, \mathbf{x}_2, \mathbf{x}_3) |\chi_1\rangle
|\chi_2 \rangle |\chi_3 \rangle,
\end{equation}
where $s_1, s_2, s_3$ encodes its spin-dependence, and
$u(\mathbf{x}_1, \mathbf{x}_2, \mathbf{x}_3)$ would be the
space-part of a spin-independent system. There is an intrinsic,
unavoidable \textit{interference effect} between the fields (much
like in the famous double-slit experiment for position) which is
lost when DIS experiments measure spin structure functions of the
``proton". The structure functions are proportional to cross
sections, which by necessity are classical quantities incapable of
encoding quantum interference. As each individual experimental
data point is a classical (non-quantum) result, structure
functions are related to incoherent sums of individual probability
distributions.

Thus, even if we (wrongly) assume the parton model to be
applicable in both cases i) and ii), S-G would result from adding
spin \textit{amplitudes} (taking full account of quantum
interference terms), while DIS would result from adding spin
\textit{probabilities} (absolute squares of amplitudes). However,
we emphasize again that in the case of S-G the parton spins are
not merely unknown, but actually \textit{undefined}. An experiment
like S-G probes the spin of the \textit{proton}, while an
experiment like DIS probes the spin of the \textit{partons} and
the final (=observed) state is not a proton at all but ``jets" of
hadrons. These two experiments are disjoint, or complementary in
the words of Bohr, and do \textit{not} describe the same physical
object.

In conclusion, we have explained why the ``proton" probed by
different experimental setups in general \textit{cannot} be
considered as the same physical object. Rather, the whole
experimental situation must be taken into account, as quantum
mechanical ``objects" and observables do not have an objective
existence unless measured. We should thus not expect to get the
same spin (1/2) for the ``proton" when measured by DIS (which
actually measures properties of the partons, not the undisturbed
proton) as when it is directly measured on the proton as a whole,
\textit{e.g.} by S-G. The ``proton" as measured by deep inelastic
scattering is a \textit{different} physical system than a
(virtually) undisturbed proton. There is no reason why spin
measurements on one should apply to the other. Especially, there
is no need for parton spins, as measured by DIS, to add up to the
spin of an undisturbed proton, just like the EMC-experiment
\cite{EMC} and its successors show. On a more pessimistic note,
DIS spin data can never directly unravel the spin of the proton
because the two are mutually incompatible. At best, it can only
serve as an indirect test of QCD by supplying asymptotic boundary
conditions to be used in future non-perturbative QCD calculations
of the proton spin. If the result of those calculations does not
come out spin-1/2, QCD is not the correct theory of strong
interactions.

\end{document}